


\font\eightrm=cmr8
\font\sixrm=cmr6

\font\ninei=cmmi9
\font\eighti=cmmi8
\font\sixi=cmmi6
\skewchar\ninei='177 \skewchar\eighti='177 \skewchar\sixi='177

\font\ninesy=cmsy9
\font\eightsy=cmsy8
\font\sixsy=cmsy6
\skewchar\ninesy='60 \skewchar\eightsy='60 \skewchar\sixsy='60

\font\eightbf=cmbx8
\font\sixbf=cmbx6

\font\ninett=cmtt9
\font\eighttt=cmtt8

\hyphenchar\tentt=-1 
\hyphenchar\ninett=-1
\hyphenchar\eighttt=-1

\font\eightsl=cmsl8

\font\eightit=cmti8



\newskip\ttglue

\def\eightpoint{\def\rm{\fam0\eightrm}%
  \textfont0=\eightrm \scriptfont0=\sixrm \scriptscriptfont0=\fiverm
  \textfont1=\eighti \scriptfont1=\sixi \scriptscriptfont1=\fivei
  \textfont2=\eightsy \scriptfont2=\sixsy \scriptscriptfont2=\fivesy
  \textfont3=\tenex \scriptfont3=\tenex \scriptscriptfont3=\tenex
  \def\it{\fam\itfam\eightit}%
  \textfont\itfam=\eightit
  \def\sl{\fam\slfam\eightsl}%
  \textfont\slfam=\eightsl
  \def\bf{\fam\bffam\eightbf}%
  \textfont\bffam=\eightbf \scriptfont\bffam=\sixbf
   \scriptscriptfont\bffam=\fivebf
  \def\tt{\fam\ttfam\eighttt}%
  \textfont\ttfam=\eighttt
  \tt \ttglue=.5em plus.25em minus.15em
  \normalbaselineskip=9pt
  \let\sc=\sixrm
  \let\big=\eightbig
  \setbox\strutbox=\hbox{\vrule height7pt depth2pt width0pt}%
  \normalbaselines\rm}

\font\twelverm=cmr10 scaled 1200    \font\twelvei=cmmi10 scaled 1200
\font\twelvesy=cmsy10 scaled 1200   \font\twelveex=cmex10 scaled 1200
\font\twelvebf=cmbx10 scaled 1200   \font\twelvesl=cmsl10 scaled 1200
\font\twelvett=cmtt10 scaled 1200   \font\twelveit=cmti10 scaled 1200


\def\twelvepoint{\normalbaselineskip=12.4pt
  \abovedisplayskip 12.4pt plus 3pt minus 9pt
  \belowdisplayskip 12.4pt plus 3pt minus 9pt
  \abovedisplayshortskip 0pt plus 3pt
  \belowdisplayshortskip 7.2pt plus 3pt minus 4pt
  \smallskipamount=3.6pt plus1.2pt minus1.2pt
  \medskipamount=7.2pt plus2.4pt minus2.4pt
  \bigskipamount=14.4pt plus4.8pt minus4.8pt
\def\rm{\fam0\twelverm}          \def\it{\fam\itfam\twelveit}%
\def\sl{\fam\slfam\twelvesl}     \def\bf{\fam\bffam\twelvebf}%
  \def\mit{\fam 1}                 \def\cal{\fam 2}%
  \def\tt{\twelvett}
  \textfont0=\twelverm   \scriptfont0=\tenrm   \scriptscriptfont0=\sevenrm
  \textfont1=\twelvei    \scriptfont1=\teni    \scriptscriptfont1=\seveni
  \textfont2=\twelvesy   \scriptfont2=\tensy   \scriptscriptfont2=\sevensy
  \textfont3=\twelveex   \scriptfont3=\twelveex  \scriptscriptfont3=\twelveex
  \textfont\itfam=\twelveit
  \textfont\slfam=\twelvesl
  \textfont\bffam=\twelvebf \scriptfont\bffam=\tenbf
  \scriptscriptfont\bffam=\sevenbf
  \normalbaselines\rm}

\def\singlespace{\baselineskip=\normalbaselineskip}
\def\doublespace{\baselineskip=\normalbaselineskip \multiply\baselineskip by 2}
\newcount\firstpageno
\firstpageno=2
\footline={\ifnum\pageno<\firstpageno{\hfil}\else{\hfil\twelverm\folio\hfil}\fi}
\let\rawfootnote=\footnote              
\def\footnote#1#2{{\rm\singlespace\parindent=0pt\rawfootnote{#1}{#2}}}
\def\dateline{\rightline{\ifcase\month\or
  January\or February\or March\or April\or May\or June\or
  July\or August\or September\or October\or November\or December\fi
  \space\number\year}}
 \def\preprintno#1{
 \rightline{\rm #1}}    

\newcount\pagenumber
\newcount\equationnumber
\newcount\citationnumber
\global\equationnumber=1
\global\citationnumber=1

\def\ifundefined#1{\expandafter\ifx\csname#1\endcsname\relax}
\def\cite#1{\ifundefined{#1} {\bf ?.?}\message{#1 not yet defined,}
\else [\csname#1\endcsname]\fi}

\def\docref#1{\ifundefined{#1} {\bf ?.?}\message{#1 not yet defined,}
\else (\csname#1\endcsname)\fi}

\def\letter{
\def\eqlabel##1{\edef##1{\the\equationnumber}}
\def\citelabel##1{\edef##1{\the\citationnumber}
\global\advance\citationnumber by1}
}

\def\no{\eqno(\the\equationnumber){\global\advance\equationnumber by1}}

\letter

\vskip -48 truept
\vskip 12 truept
\def\frac#1#2{{\textstyle{#1 \over #2}}}
\def\square{\kern1pt\vbox{\hrule height 1.2pt\hbox{\vrule width 1.2pt\hskip 3pt
   \vbox{\vskip 6pt}\hskip 3pt\vrule width 0.6pt}\hrule height 0.6pt}\kern1pt}
\hsize=6.5truein
\hoffset=.1truein
\vsize=8.7truein
\voffset=.05truein
\parskip=\medskipamount
\twelvepoint            
\doublespace            
\overfullrule=0pt       

\vskip -0.4truein
\preprintno{DIAS-STP-95-24}
\dateline
\centerline{\bf Modular Invariance of Finite Size Corrections and a Vortex
Critical Phase.}
\vskip\baselineskip
\centerline{by}
\vskip\baselineskip
\centerline{Charles Nash$^{*,{\dag}}$ and Denjoe O' Connor$^{\dag}$}
\par\vskip\baselineskip
\noindent
$^*$Department of Mathematical Physics\hfill
$^{\dag}$School of Theoretical Physics
\vskip-.5\baselineskip\noindent
St. Patrick's College\hfill              Dublin Institute for Advanced
Studies
\par\vskip-.5\baselineskip\noindent
Maynooth\hfill                {\it and}\hfill  10 Burlington Road
\par\vskip-.5\baselineskip\noindent
Ireland\hfill                            Dublin 4
\par\vskip-.5\baselineskip\noindent
\null\hfill                              Ireland
\par\vskip1.5\baselineskip
\noindent{\bf Abstract: }
We analyze a continuous spin Gaussian model
on a toroidal triangular lattice with periods
$L_0$ and $L_1$  where the spins
carry a representation of the fundamental group of the torus labeled
by phases $u_0$ and $u_1$. We find
the {\it exact finite size and lattice corrections},
to the partition function $Z$, for arbitrary mass $m$ and phases $u_i$.
Summing $Z^{-1/2}$ over phases gives the corresponding result for the Ising
model.
The limits $m\rightarrow0$ and $u_i\rightarrow0$  do not commute.
With $m=0$ the model exhibits a {\it vortex critical phase}
when at least one of the $u_i$ is non-zero.
In the continuum or scaling limit, for arbitrary $m$, the finite size
corrections to
$-\ln Z$ are {\it modular invariant} and for the critical phase are given by
elliptic theta functions. In the cylinder limit $L_1\rightarrow\infty$
the ``cylinder charge'' $c(u_0,m^2L_0^2)$ is a non-monotonic function of
$m$ that ranges from $2(1+6u_0(u_0-1))$ for $m=0$ to zero for
$m\rightarrow\infty$.
\vfill\eject
\noindent
Finite size effects are an intrinsic feature of laboratory experiments
which probe the neighbourhood of a continuous phase transition.
The scaling properties of these corrections to the
infinite (bulk) system behaviour play an increasingly
important role in computer simulations and our theoretical understanding
of the critical regime of statistical
systems\citelabel{\BarbrCardyBook}\cite{BarbrCardyBook}.
The interplay of these effects with other aspects of the system
can give rise to crossover from one characteristic behaviour to
another\citelabel{\EnvfRG}\cite{EnvfRG}.
It is possible to make real progress in the detailed analysis
of such situations in two dimensions and for this reason
two dimensional models have attracted much interest in recent years.
Many of their properties are expressible in terms of those of generalized
Gaussian models\citelabel{\DotsenkoFateev} \cite{DotsenkoFateev}
and with minor modification the results
of this note can be adapted to a wide class of other models.
\par
Pure finite size effects can be isolated from those due to the
presence of a boundary by ensuring that the connectivity of
the underlying lattice is such that it has no boundary.
One can avoid other complications, such as local curvature
of the lattice, while retaining the finite size effects,
by considering a {\it flat torus}. Furthermore, in two dimensions,
the {\it only} zero curvature finite volume manifold without boundary
is the {\it flat torus}. This torus can be conveniently thought of as
a parallelogram with opposite sides identified.
We take the sides to be of length
$L_0$ and $L_1$,  with $L_0$ at  an  angle $\theta$
to $L_1$, cf. fig. 1. Simple geometry then means that the point $(x,y)$
is thereby identified with the point $(x+aL_1\cos\theta+b
L_0,y+aL_1\sin\theta)$
where $a$ and $b$ are integers.
\par
We consider a triangular lattice composed of similar triangles, pairs of which
form parallelograms,
cf. fig. 1.  The basic triangles have two sides of lengths $a_0$ and $a_1$ with
an angle $\theta$ between them. The complete lattice forms our torus, $T^2$,
and consists of $K_0K_1$ sites and $2K_0K_1$ triangles,
forming a parallelogram of sides $L_0=K_0a_0$ and $L_1=K_1a_1$.
\par
We take a complex continuous spin
variable $\varphi(x,y)$ on $T^2$ but will {\it not} demand that $\varphi$ be
periodic,
though the underlying lattice is,  rather it acquires a phase on
being transported around one of the cycles of the torus. In general we
require that
\eqlabel{\bdryconds}
$$\varphi(x+aL_1\cos\theta+bL_0,y+aL_1\sin\theta)=e^{2\pi
i(au_1+bu_0)}\varphi(x,y),\quad a,b\in{\bf Z}\no$$
For $u_0$ and $u_1$ neither integer or half integer the spin variable is
necessarily either complex or
real with two components: From the two component viewpoint these boundary
conditions
can be thought of as corresponding to vortices winding around the periods of
the
torus. From a mathematical standpoint  $\varphi$ is a section
of a bundle ${\cal L}$ over the torus $T^2$.
\par
Labeling the lattice sites by $(k_0,k_1)\equiv{\bf k}=k_0+(k_1-1)K_0$
with $k_i=1,\dots,K_i$, the energy of a configuration is given by
\eqlabel{\quadform}
$$\eqalign{{\cal E}_{\cal L}[T^2,\varphi^*,\varphi]&={1\over2}\sum_{{\bf k}{\bf
k}^\prime}
\sqrt{g}\varphi^{*}({\bf k})\left(\Delta({\bf k},{\bf k}^\prime)
+m^2\delta_{{\bf k},{\bf k^\prime}}\right)\varphi({\bf k}^\prime)\cr
}\no$$
where $\sqrt{g}=a_0a_1\sin\theta$ and
$\delta_{{\bf k},{\bf k}^\prime}$ is the Kronecker delta.
$\Delta$ is a  $K_0K_1\times K_0K_1$ {\it symmetric} matrix,
all entries of  which are determined by nearest neighbour
interactions, so that each spin
interacts with six neighbouring spins.  Explicitly
the only non-zero elements are those given below (and their transposes): we
have
$$\eqalign{
\Delta\{(k_0,k_1),(k_0+1,k_1)\}&=-\alpha=-{1\over\sin^2\theta}
\left({1\over a_0^2}-{\cos\theta\over a_0a_1}\right)\cr
\Delta\{(k_0,k_1),(k_0,k_1+1)\}&=-\beta=-{1\over\sin^2\theta}
\left({1\over a_1^2}-{\cos\theta\over a_0a_1}\right)\cr
\Delta\{(k_0+1,k_1),(k_0,k_1+1)\}&=-\gamma=-{1\over\sin^2\theta}\left({\cos\theta\over a_0a_1}\right)\cr
\Delta\{(k_0,k_1),(k_0,k_1)\}&=2\sigma,\quad\hbox{where
}\sigma=\alpha+\beta+\gamma\cr
\cr}
\no$$
We actually solve the model for general nearest neighbour couplings,
$\alpha$, $\beta$ and $\gamma$, in which case
$g=(\alpha\beta+\beta\gamma+\gamma\alpha)^{-1}$.
Despite the simplicity of the model
we will see that it has a surprisingly rich structure.
\par
The  eigenfunctions of $\Delta$ that satisfy \docref{bdryconds}
are the discrete Fourier basis
\eqlabel{\lateigenfns}
$$e_{mn}(k_0,k_1)={1\over\sqrt{K_0 K_1}}
\exp\left[2\pi i\left\{(m+u_0){k_0\over K_0}+(n+u_1){k_1\over
K_1}\right\}\right],\quad\hbox{with }\cases{m=1,\dots K_0\cr
                                            n=1,\dots K_1\cr}\no$$
orthogonality of which is given by
\eqlabel{\orthogonality}
$$\sum_{k_0=1}^{K_0}\sum_{k_0=1}^{K_1}e_{mn}^*(k_0,k_1)e_{m^\prime
n^\prime}(k_0,k_1)=\delta_{m,m^\prime}\delta_{n,n^\prime}\no$$
while the corresponding sum over  $m$ and $n$ gives completeness.
The eigenvalues $\lambda_{n_0n_1}$ of the matrix ${1\over2}(\Delta+m^2)$
are then
\eqlabel{\lateigenvals}
$$\eqalign{\lambda_{n_0n_1}
&=\delta-\alpha\cos\left(x_{n_0}\right)-\beta\cos\left(x_{n_1}\right)
-\gamma\cos\left(x_{n_0}-x_{n_1}\right)
\cr
 \hbox{where }
x_{n_i}&={2\pi(n_i+u_i)\over K_i},\;i=0,1
\qquad\hbox{ and}\qquad \delta=\sigma+{m^2\over2}
\cr}
\no$$
\par
We are interested in the partition function
$$Z(T^2,{\cal L},m)=\int\left[\prod d\varphi^* d\varphi\right]
e^{-{1\over k_BT}{\cal E}_{\cal L}[T^2,\varphi^*,\varphi]}
=\prod_{(n_0,n_1)}^{(K_0,K_1)}\left\{{\pi k_BT\over
\sqrt{g}\lambda_{n_0n_1}}\right\}
\no$$
The free energy is then given by
$F=k_BTW$ where $W=-\ln Z$.
\par
To perform the sums we note the rearrangement of the eigenvalues
$$\eqalign{&\lambda_{n_0n_1}
=\delta-\alpha\cos x_{n_0}-\vert \beta_{n_0}\vert\cos(x_{n_1}-\theta_{n_0})
\quad\hbox{where}\quad
\beta_{n_0}=\beta+\gamma e^{ix_{n_0}} =\vert \beta_{n_0}\vert
e^{i\theta_{n_0}}\cr
}\no$$
and use the basic non-trivial identity\citelabel{\Anatoly}\cite{Anatoly}
\eqlabel{\identity}
$$\eqalign{
{\sum_{n=0}^{K-1}}
\ln\left[z-\cos x_n\right]
&=K\int_{0}^{{\pi\over2}}{d\nu\over\pi}\ln\left[z^2-\cos^2\nu\right]
+\ln\left\vert 1-{\left(z-\sqrt{z^2-1}\right)}^Ke^{2\pi i u}\right\vert^2\cr
}\no$$
where $z\ge1$ and real. In any case the sum over $n_1$ followed by that
over $n_0$ yields:
\eqlabel{\wtotal}
$$W=K_0K_1W_B+W_F\no$$
where
\eqlabel{\wbulk}
$$W_B=
\int_{-\pi}^{\pi}{d\nu_1\over 2\pi}\int_{-\pi}^{\pi}{d\nu_2\over 2\pi}
\ln\left[{\sqrt{g}\over\pi k_BT}\left\{
\delta-\alpha\cos(\nu_1)-\beta\cos(\nu_2)-\gamma\cos(\nu_1-\nu_2)
\right\}\right]\no$$
and
\eqlabel{\wf}
$$\eqalign{W_F=&K_1\sum_{\epsilon=\pm}\int_{0}^{\pi\over2}{d\nu\over\pi}
\ln\left\vert1-{\left(q_{\epsilon}-\sqrt{q_{\epsilon}^2-1}\right)}^{K_0}e^{2\pi
i u_0}\right\vert^2\cr
&\qquad+\sum_{n_0=-\left[{K_0\over2}\right]}^{\left[{K_0-1\over2}\right]}
\ln\left\vert1-e^{-K_1v_{n_0}}e^{2\pi iu_1}\right\vert^2}\no$$
here $[x]$ denotes the integer part of $x$.
The definitions of the various variables in \docref{wf} above are as follows:
$$\eqalign{ q_{\pm}&=q\pm\sqrt{q^2-p}\qquad
q={\alpha\delta+\beta\gamma\cos^2\nu\over \alpha^2},\qquad
p={\delta^2-(\beta^2+\gamma^2)\cos^2\nu\over \alpha^2}\cr
\cr}\no$$
and
$$v_{n_0}=-\ln\left[z_{n_0}-\sqrt{z_{n_0}^2-1}\right]+i\theta_{n_0},\quad\hbox{
where}\quad
z_{n_0}={\delta -\alpha\cos x_{n_0}\over 2\vert \beta_{n_0}\vert}\no$$
\par
Note that $W_B$ gives the free energy per lattice site in the thermodynamic
limit,
i.e.
$$\lim_{K_0,K_1\rightarrow\infty}{W \over K_0K_1}=W_B\no$$
hence, due to the fact that we are dealing with a finite
rather than an infinite lattice,
$W_F$ gives the {\it complete finite size correction} to the
bulk lattice behaviour.
\par
If instead we perform the sum over $n_0$  before that over $n_1$
we find the alternative composition $W=K_0K_1W_B+\tilde W_F$.
The two apparently different expressions $W_F$ and
$\tilde W_F$ are, obviously, equal;
they can be transformed into one another
by the interchange of $\alpha$ with $\beta$ and the
subscripts $0$ and $1$.
\par
The limit of interest to us is the {\it continuum} or {\it scaling}  limit.
This is a constrained thermodynamic limit
achieved by taking  $K_0,K_1\rightarrow\infty$ while keeping fixed
$L_i=K_ia_i$, $\theta$, $m^2$ and the ratio $k={K_1\over K_0}$.
The asymptotic form of
$W_B$ in this scaling limit is given by
$$\eqalign{W_B K_0K_1
&=K_0K_1\Lambda_B-{Vm^2\over4\pi}\left\{\ln[K_0K_1]-2\rho\right\}
-{Vm^2\over4\pi}(\ln[{m^2V\over 4\pi^2}]-1)+\cdots\cr
}\no$$
Here $\Lambda_B$ is the value of $W_B$ at $m=0$ and
$$\eqalign{\rho\equiv\rho(\alpha, \beta, \gamma)&=\int_{0}^{\pi}d\nu
\left[{1\over\sin\nu\sqrt{1+g\alpha^2\sin^2\nu}}-{1\over\nu}\right]
-{1\over2}\ln\left[\sqrt{g}(\beta+\gamma)\right]\cr
}\no$$
Despite its appearance, $\rho(\alpha,\beta,\gamma)$  is
{\it symmetric} under interchange of
$\alpha$ and $\beta$.  Both $\Lambda_B$ and
$\rho$ depend on $L_0$, $L_1$, $\theta$ and $k$, or equivalently they
depend on the geometry of the lattice triangle.
\par
We find the limiting behaviour of
$W_F$ is given by\footnote{$^{(a)}$}{\eightpoint
In extracting the limit, the expansions
$\ln q_-=-{\sqrt{g}\over2}(\beta+\gamma)^2(\nu^2+{g m^2 \over \beta+ \gamma})
+\cdots$
and \hfill\break\line{
$\eqalign{v_{n_0}
&=\sqrt{{1\over g (\beta+\gamma)\vert \beta+\gamma\vert}x_{n_0}^2
+{m^2\over\vert \beta+\gamma\vert}}+i {\gamma\over \beta+\gamma}x_{n_0}
+\cdots  \cr}$
for small $\nu$ and $m^2$ are useful.\hfill}
}
\eqlabel{\wflimit}
$$\eqalign{\Gamma_F =&-{\pi\tau_1\over6} c(u_0,{m^2V\over\tau_1})
+\sum_{n=-\infty}^{\infty}
\ln\left\vert1-e^{-2\pi \tau_1\sqrt{{(n+u_0)}^2+{m^2V\over4\pi^2\tau_1}}+2\pi
i\{u_1-\tau_0(n+u_0)\}}\right\vert^2
\cr}\no$$
where
$$\tau_0={k\over\sqrt{g}\vert\beta+\gamma\vert},\qquad
\tau_1={k\gamma\over\beta+\gamma}\quad\hbox{ and} \qquad V=K_0K_1\sqrt{g}\no$$
The function $c(u,x)$ that appears in \docref{wflimit} is given by
\eqlabel{\cdef}
$$\int_{-\infty}^{\infty}{dp\over2\pi}
\ln\left\vert1-e^{-\sqrt{p^2+x}+2\pi i u}\right\vert^2=
-{c(u,x)\pi\over6}\no$$
Corrections to \docref{wflimit} which vanish in the limit are also easily
obtainable from \docref{wf}.
In terms of the torus geometry  described above we have
$\tau_0={L_1\over L_0}\cos\theta$, $\tau_1={L_1\over L_0}\sin\theta$
and $V=L_0L_1\sin\theta$.

Certain special cases of \docref{wflimit} arise in the case of an Ising model
on a
torus (see \citelabel{\FerdFisher}\cite{FerdFisher} where the case $\tau_0=0$
was studied).  This can be understood from the equivalence of the Ising model
to
a dimer model on a decorated lattice where similar determinants
arise\citelabel{\KastelynFisher}\cite{KastelynFisher}.
{}From this equivalence, exhibiting the phase dependence
of $W_F$ by $W_F(u_0,u_1)$, with $\alpha$, $\beta$, $\gamma$ and $\delta$
obtained
from \citelabel{\Stephenson}\cite{Stephenson}, we have
\eqlabel{\Zising}
$$Z^{Ising}={1\over2}e^{-W_{B}^{Ising}}\left\{
\mp
e^{{1\over2}W_F(0,0)}+e^{{1\over2}W_F(0,{1\over2})}+e^{{1\over2}W_F({1\over2},0)}
+e^{{1\over2}W_F({1\over2},{1\over2})}\right\}\no$$
for ferromagnetic couplings, with $+$ referring to $T<T_c$ and $-$ to $T>T_c$.
In the scaling limit $W_F\rightarrow\Gamma_F$.
Our results therefore incorporate the complete
lattice and finite size corrections for the Ising model on a triangular
lattice.
With a similar equivalence
our results can easily be translated to give the general
result for other models.
\par
$\Gamma_F$ is {\it invariant} under the transformations
$$\eqalign{\hbox{(i)}\qquad &u_0\mapsto u_0
\qquad\qquad u_1\mapsto u_0+u_1\qquad
\tau_0\mapsto\tau_0+1\qquad \tau_1\mapsto\tau_1\cr
\hbox{(ii)}\qquad&u_0\mapsto u_1\qquad\qquad u_1\mapsto -u_0
\qquad\quad\tau_0\mapsto -
{\tau_0\over\vert\tau\vert^2}\qquad
\tau_1\mapsto{\tau_1\over\vert\tau\vert^2}\cr}$$
where $\tau=\tau_0+i\tau_1$. Invariance under (i) is obvious from
\docref{wflimit} and records
the action
of the generator `$T$' of $SL(2,{\bf Z})/{\bf Z}_2$ while (ii) interchanges
$\Gamma_F$ with $\tilde \Gamma_F$ this latter being obtained by performing the
sums leading to $\Gamma_F$ in the opposite order. These two expressions are
{\it equal} and
(ii) records the action of the generator `$S$' of  the modular group $SL(2,{\bf
Z})/{\bf Z}_2$.
The volume $V$ is invariant under the action these generators.
But $S$ and $T$ generate the whole modular group  $SL(2,{\bf Z})/{\bf Z}_2$ and
so $\Gamma_F$  and $V$ are invariant under the
action of any element of $SL(2,{\bf Z})/{\bf Z}_2$.
Hence we can conclude that {\it in the scaling limit
the complete finite size corrections to the bulk free energy
are invariant under the entire modular group}.
\par
If $M$ denotes an arbitrary element of the modular group,
under which $\tau$ undergoes the well known transformation
$\tau\mapsto(a\tau+b)/(c\tau+d)$,  we have that $\Gamma_F$
is invariant under the replacement
$$\eqalign{&u_1\mapsto u_1[M]=a u_1+b u_0 \qquad
u_0\mapsto u_0[M]=c u_1+ du_0\cr
&L_1\mapsto L_1[M]=a L_1+b L_0e^{-i\theta}\qquad L_0\mapsto
L_0[M]=cL_1e^{i\theta}+dL_0\cr}\no$$
where
$$M=\pmatrix{a&b\cr
c&d\cr}\in SL(2,{\bf Z})/{\bf Z}_2,\quad a,b,c,d\in{\bf Z},\; ab-cd=1\no$$
We note that the sum over the four terms in \docref{Zising} gives the finite
size
correction $Z_F^{Ising}$ as a modular invariant function of $\tau_0$, $\tau_1$
and $m^2V$.
It is clear from our construction that summing over the phases $u_i$ which form
an
orbit of $SL(2,{\bf Z})/{\bf Z_2}$ on the space of phases (the Picard Variety
\citelabel{\RaySinger}\cite{RaySinger}) will allow one to construct similar
phase independent modular invariant partition functions,
such as arise in other conformal field theories
cf.\citelabel{\ItsonDroufe}\cite{ItsonDroufe}
and references therein.
\par
Taking $m\rightarrow 0$ in \docref{wflimit} we find the critical phase
limit of $\Gamma_F$ is given by
$$\eqalign{\Gamma_F=&-{\pi\tau_1\over6}c(u_0)+\sum_{n=-\infty}^{\infty}
\ln\left\vert1-e^{2\pi i\left\{\tau \vert n\vert -\epsilon_n (u_1-\tau
u_0)\right\}}\right\vert^2,\quad \epsilon_n={n+u_0\over \vert n+u_0\vert}\cr
=&\ln\left\vert{e^{\pi i u_0^2\tau}\vartheta_1(u_1-\tau
u_0\vert\tau)\over\eta(\tau)}\right\vert^2\cr}\no$$
where $c(u)=2\{1-6u(1-u)\}$,  $\vartheta_1$ is a Jacobi theta function
and $\eta$ is the Dedekind eta function---the
modular transformation properties of
$\vartheta_1$ and $\eta$  are well known and
allow an independent check
of the modular invariance in this limit.
\par
A further limit of interest is the geometric one obtained by taking
 $L_1\rightarrow\infty$; it corresponds to
a cylindrical geometry. In this
limit $\Gamma_F/V$ reduces to the expression $\gamma^{cylinder}$
where
\eqlabel{\cylin}
$$\gamma^{cylinder}=-{\pi\over 6L_0^2}c(u_0,m^2L_0^2)\no$$
(The analogous limit,  where $L_0\rightarrow\infty$,
 replaces $u_0$ and $L_0$
by $u_1$ and $L_1$ respectively.) When $u_0$ and $m$ are both zero
the ``cylinder charge''   $c(u_0,m^2L_0^2)$ reduces to
the central charge of the model;  in our case $c=2$.
The cylinder charge should not be confused with the
Zamolodchikov $c$-function \citelabel{\Zamolodchikov}\cite{Zamolodchikov}
for this model, the two functions have different dependencies on $m$, the
latter
for example being a monotonic function of $m$.
We plot $c(u,x)$ in fig. 2 for
different values of $u$ as a function of $x$.
It exhibits the crossover to $c=0$ at large values of $x$,
and is clearly {\it not} a monotonic function of $x$.
{}From \docref{Zising} we see that
$$\gamma^{cylinder}_{Ising}={\pi\over12L_0^2}c({1\over2},x)\no$$
Comparison with \docref{cylin} means that the cylinder
charge for the Ising model is $-{1\over2}c({1\over2},x)$, which for $x=0$
gives the usual central charge
$c={1\over2}$\citelabel{\BloteCardyNightengaleAffleck}\cite{BloteCardyNightengaleAffleck}.
\par
The limits $m\rightarrow0$ and $u_0,u_1\rightarrow0$ for $\Gamma_F$ do not
commute.
To see this we expand $\Gamma_F$ \docref{wflimit} for small $m$ and $u_0,u_1$
obtaining
\eqlabel{\asymptotic}
$$\Gamma_F =\ln\left[(2\pi)^2\vert u_1-\tau u_0\vert^2+\tau_1 m^2V\right]
+2\ln\vert\eta(\tau)\vert^2+\cdots\no$$
and this expression clearly can tend to the distinct
logarithmically singular expressions
$\ln\left\vert u_1-\tau u_0\right\vert^2+2\ln\vert\eta(\tau)\vert^2$ and
$\ln[\tau_1 m^2V]+2\ln\vert\eta(\tau)\vert^2$ depending on the
order in which the limits are taken. However both limits and indeed
\docref{asymptotic} itself are {\it modular invariant}.
\par
The continuum version of the model studied here is
described by the  Hamiltonian
\eqlabel{\genmetricaction}
$$I[\varphi,L_0,L_1,\theta]={1\over 2}\int_{T^2}\sqrt{g}
\left[\partial_{\mu}\varphi^*g^{\mu\nu}\partial_{\nu}\varphi+m^2\varphi^*\varphi\right]\no$$
It is a simple task to obtain the corresponding continuum partition function
by the method of $\zeta$-function regularisation. This procedure yields the
finite size contribution $\Gamma_f$ of \docref{wflimit}
but the bulk term differs significantly being now
$-{Vm^2\over4\pi}(\ln[(m/2\pi\mu)^2]-1)$ where $\mu$ is
an arbitrary undetermined scale.  Analogues of this continuum model
were discussed in \cite{ItsonDroufe} where the
phases were taken to be rational numbers.  Their expressions
unfortunately contain errors and they did not succeed in
separating the bulk and finite size contributions.
\par
The model we have studied above serves as a useful starting point for a
perturbative treatment of the approach to the Kosterlitz-Thouless
phase\citelabel{\KostThou}\cite{KostThou} in a $\vert\varphi\vert^4$
theory and thus captures the neighbourhood of the critical point of
the $XY$ model as the critical point is approached from
the disordered phase. That the boundary conditions
capture the essential features of a vortex phase can be
seen by considering \docref{bdryconds} in  the cylinder limit and
mapping the cylinder to the plane using the conformal map
$y+i{2\pi x\over L_0}=\ln z$.
The charge of the vortex at the origin is then given by $u_0$ and its
existence gives rise to an Aharanov-Bohm effect for transport around
the origin.
\par
In summary, the finite size corrections to the free energy are {\it modular
invariant.}
This conclusion extends to the entire scaling neighborhood of the critical
phase.
We use our results to give expressions for the complete lattice and
finite size corrections for the two dimensional Ising model on a
triangular lattice via its equivalence to a sum over Pfaffians.
Modular invariance also extends to
models in more than two dimensions when the geometry giving rise to finite size
effects contains a flat torus. For a three dimensional
cylindrical geometry with toroidal cross-section the result can be obtained
from
$\Gamma_f$ by replacing $m^2$ with $m^2+q^2$ and integrating the resulting
expression over $q$. One can understand the origin of modular invariance in
general
as the residual freedom to reparametrize coordinates, in the continuum limit,
while retaining flat toroidal geometry.
\par
In the two dimensional case the limiting finite size corrections
at the critical phase are expressible
in terms of classical elliptic functions. Infinitesimally small values of the
phases
$u_i$ lead to logarithmically  divergent contributions to the free energy. This
implies that the free energy needed to create a vortex
becomes infinite for an infinitely large lattice.  In general the model has a
surprisingly rich structure of non-commuting limits. For example the
limits of approaching the critical phase and that of sending the $u_i$ to
zero do not commute.
\par
{\bf Acknowledgment:} We are grateful to Paul Upton for helpful conversations
and
his careful reading of the manuscript.
\par\vfill\eject
\input epsf
\epsfxsize=0.6\hsize
\centerline{\bf The triangulated torus}
\par\vskip \baselineskip
\centerline{\epsffile{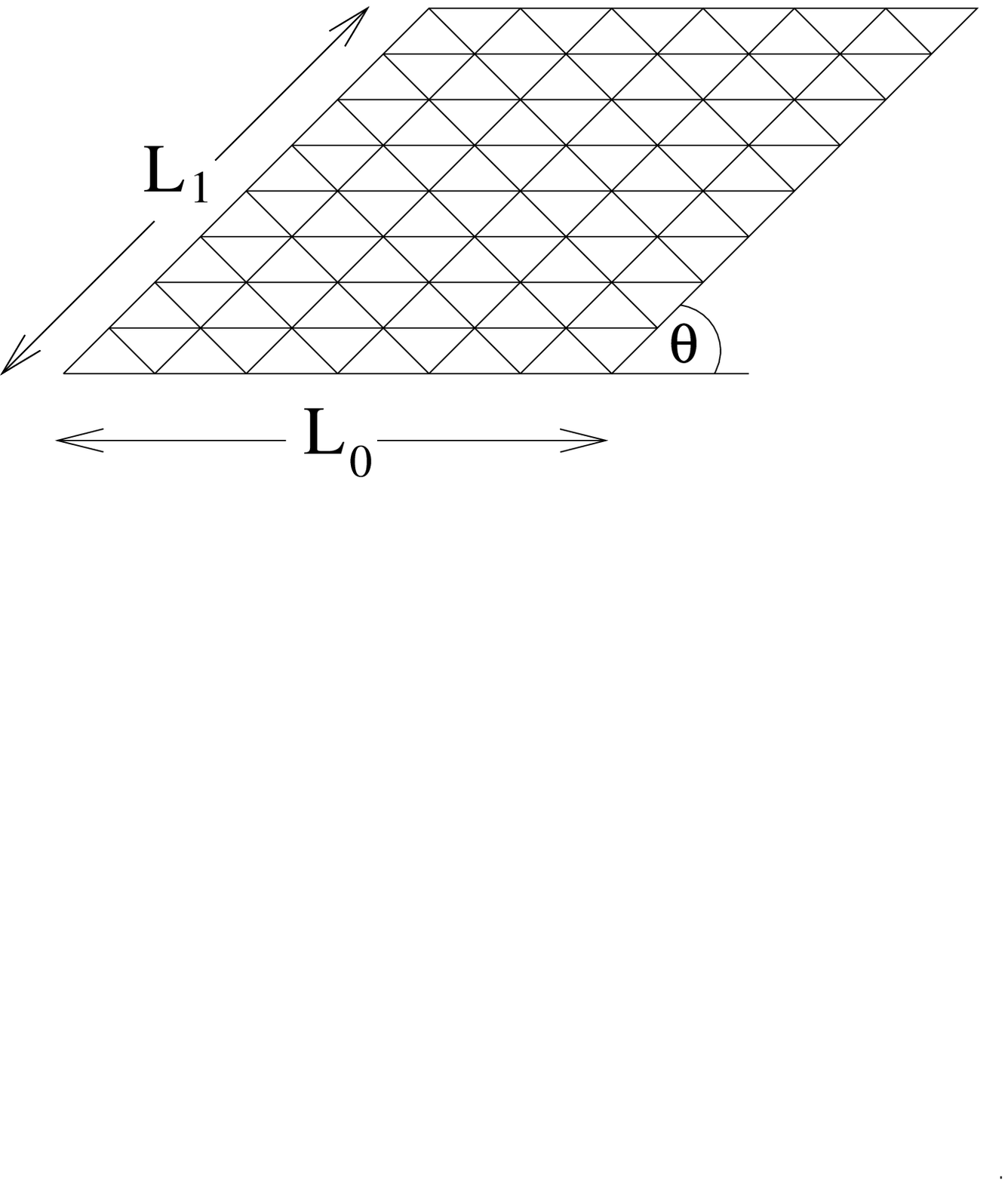}}
\par\vskip-0.25\vsize
\par\vskip-1.5\baselineskip
\centerline{\bf Fig. 1.}
\par\vskip \baselineskip
\centerline{\bf The `cylinder charge'  function $c(u,x)$ for various $u$}
\epsfxsize=0.6\hsize
\par\vskip-0.1\vsize
\centerline{\epsffile{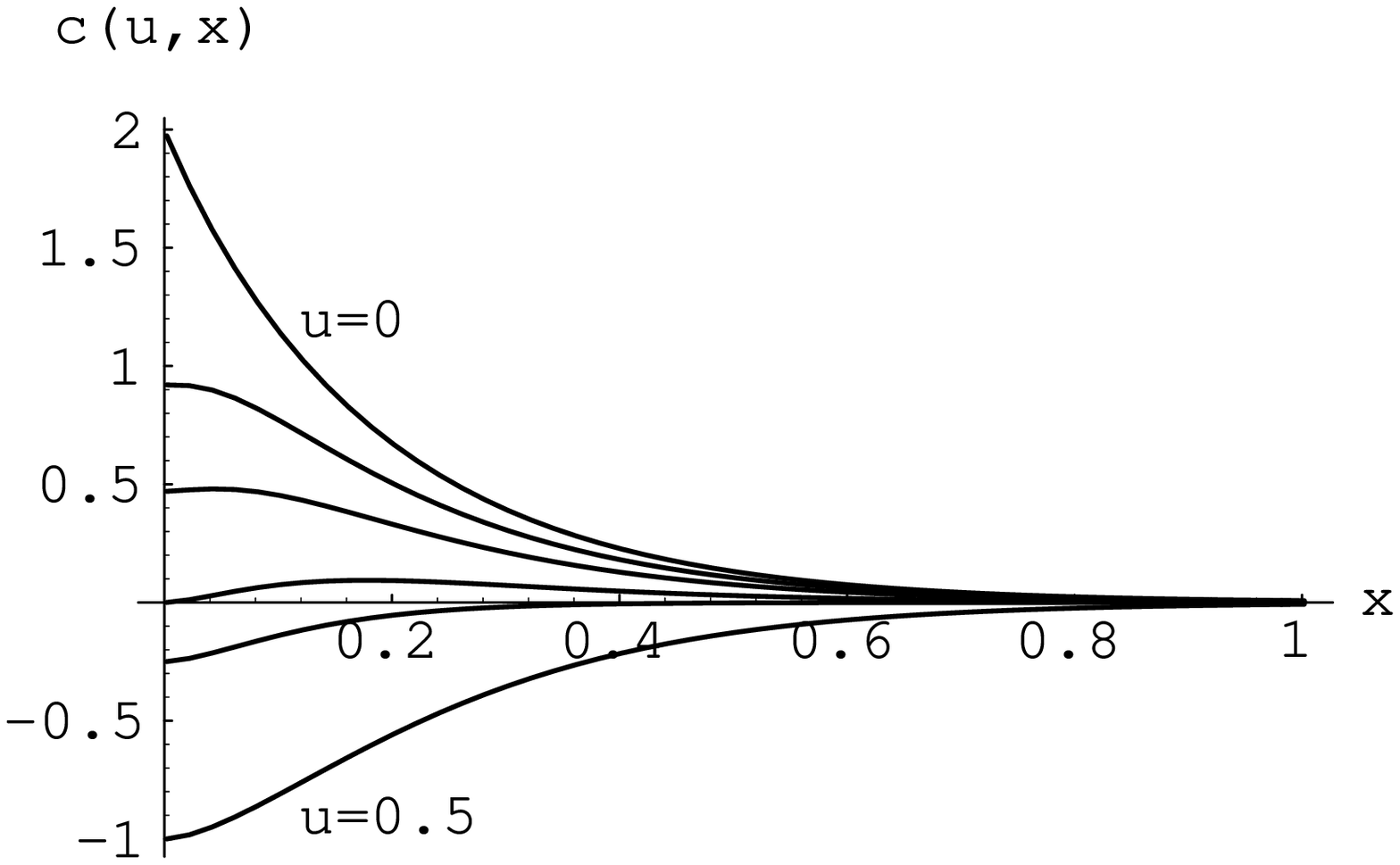}}
\par\vskip-1.5\baselineskip
\centerline{\bf Fig. 2.}
\vfill\eject
\noindent
{\bf References: }
\item{\cite{BarbrCardyBook}}M. N. Barber, in Phase Transitions and Critical
Phenomena,
vol.8, eds.  C. Domb and J. L. Lebowitz
(Academic Press, London 1983).
 {\it Finite Size Scaling} CPSC vol.2,
ed. J. L. Cardy (North Holland, 1988).

\item{\cite{EnvfRG}} D.\ O'Connor and C.R.\ Stephens,
  {\it Int.\ J.\ Mod.\ Phys.}\ {\bf A9}, 2805 (1994).

\item{\cite{DotsenkoFateev}} Vl. S. Dotsenko and V. A. Fateev,
{\it Nucl. Phys.} {\bf B240} 312 (1984), and {\bf B251}, 691 (1985).

\item{\cite{Anatoly}}This is a generalization of an identity in
A. E. Patrick,  {\it J. Stat. Phys.} {\bf 72}, 665 (1993).

\item{\cite{FerdFisher}} A. E. Ferdinand and M. E. Fisher,
{\it Phys. Rev.} {\bf 185} 832 (1969).

\item{\cite{KastelynFisher}}P.W. Kastelyn, {\it J. Math. Phys.} {\bf 4}, 287
(1963).
M. E. Fisher, {\it J. Math. Phys.} {\bf 7}, 1776 (1966).

\item{\cite{Stephenson}}J. Stephenson, {\it J. Math. Phys.} {\bf 8}, 1009
(1964).

\item{\cite{Zamolodchikov}}A. B. Zamolodchikov, {\it Pis'ma Zh. Eksp. Teor.
Fiz.}
{\bf 43}, 565 (1986) [{\it JETP Lett.} {\bf 43}, 730 (1986).

\item{\cite{BloteCardyNightengaleAffleck}}
H. W. J. Bl\"ote,  J. L. Cardy
and M. P. Nightingale, {\it Phys. Rev. Lett.} {\bf 56}, 742 (1986).
I. Affleck,  {\it Phys. Rev. Lett.} {\bf 56}, 746 (1986).

\item{\cite{RaySinger}}D. B. Ray and I.  M. Singer, {\it Ann. Math.} {\bf 98},
154 (1973).

\item{\cite{ItsonDroufe}}C.  Itzykson and J. Drouffe,
{\it Statistical Field Theory} vol 2, C.U.P. (1989).

\item{\cite{KostThou}} J. M. Kosterlitz and D. J. Thouless,
{\it J. Phys.} {\bf C 6}, 1181 (1973).

\bye